\documentclass[aps,prd,onecolumn,superscriptaddress,showpacs,floatfix,nofootinbib]{revtex4}
\usepackage{psfig}
\usepackage{comment}
\usepackage{graphicx}
\usepackage[dvips]{color}
\usepackage{bm}

\newcommand{\met}{ $/\!\!\!\!E_{T}$} 

\begin{document}

\vspace{2cm}
\preprint{CDF/PUB/EXOTIC/PUBLIC/10921}
\preprint{Draft version 1.0}


%
%

\title{Standard Model Higgs Searches at the Tevatron\footnote{This is
based on the talk I gave at the phenomenology 2012 symposium, 
7-9 May 2012, University of Pittsburgh.}}

\author{Weiming Yao (LBNL)\\ (For the CDF and D0 Collaborations)}




\begin{abstract}
We present the results of direct searches for the standard model Higgs boson 
at the Tevatron. Results are derived from the
complete Tevatron Run II dataset, with a measured integrated luminosity of
10 fb$^{-1}$ of proton-antiproton data. The searches are performed for assumed
Higgs masses between 90 and 200 GeV/c$^2$. We observe an excess of events 
in the data compared with the background predictions, which is most 
significant in the mass range between 115 and 135 GeV/c$^2$, consistent with 
the Higgs-like particle recently observed by ATLAS and CMS. The largest local 
significance is 2.7 standard deviations, corresponding to a global significance
of 2.2 standard deviations. We also combine separate searches for 
$H\rightarrow b\bar b$ and $H\rightarrow W^+W^-$, and find that the excess
is concentrated in the $H\rightarrow b\bar b$ channel, although the results 
in the $H\rightarrow W^+W^-$ channel are still consistent with the possible 
presence of a low-mass Higgs boson. 
 

\end{abstract}


\maketitle

\section{Introduction}	
The Higgs boson was hypothesised as a remnant of the Higgs field that 
was responsible for the electroweak symmetry breaking about a half century 
ago~\cite{higgs}. Understanding the mechanism for electroweak symmetry 
breaking, especially by testing for the presence or absence of the standard 
model (SM) Higgs boson, has been a major goal of particle physics 
and a central part of the Fermilab Tevatron physics program. Both CDF and D0 
collaborations have performed new combinations of multiple direct searches for 
the standard model Higgs boson~\cite{tevcomb}. The new searches include 
more data, additional 
channels, and improved analyses techniques compared to previous analysis. 
Results are derived from the complete Tevatron Run II dataset, with a measured 
integrated luminosity of 10 fb$^{-1}$ of proton-antiproton data. 
The searches are performed for assumed Higgs masses 
between 90 and 200 GeV/c$^2$.

The global fit of the electroweak precision data, including recent 
top-quark and $W$ boson mass measurements from the Tevatron~\cite{mtop,mw},
constrains the Higgs mass $m_H$ to be less than 152 GeV/c$^2$ at the 95\% 
conference level (CL)~\cite{ewfit}.
The direct searches from LEP~\cite{lep}, Tevatron~\cite{tevcomb}, and 
LHC results~\cite{cms,atlas} set the Higgs mass between 
116.6 and 119.4 GeV/c$^2$ or between 122.1 and 127 GeV/c$^2$ 
at the 95\% CL. Recently both LHC experiments~\cite{discoverycms, 
discoveryatlas} observed local excesses above 
the background expectations for a Higgs boson mass of approximately 125 
GeV/c$^2$. Much of the power of 
the LHC searches comes 
from $gg\rightarrow H$ production and Higgs boson decays to $\gamma\gamma$, 
$W^+W^-$, and $Z^+Z^-$, which probe the couplings of the Higgs boson to other 
bosons. In the allowed mass range, the Tevatron experiments are particularly 
sensitive to the association production of the Higgs boson with a weak vector 
boson in the $b\bar b$ channel, which probes the Higgs boson's couplings to
$b$ quarks.

The Tevatron collider produces proton and anti-proton collision at the center
mass of 1.96 TeV with a record luminosity of 4.3~$10^{32}$~cm$^{-2}$s$^{-1}$. 
The Tevatron has delivered close to 12 fb$^{-1}$ to each experiment before 
the shutdown on 30 September 2011 after 28 successful years running. 
Both CDF and 
D0 detectors are the general-purpose detectors, which provide excellent 
tracking, lepton identification, jets finding, and missing transverse energy 
 (\met) detections. The details can be found elsewhere~\cite{cdf,d0}. 

\section{Higgs Production and Decays} 

\begin{figure}[htpb]
\centerline{\psfig{file=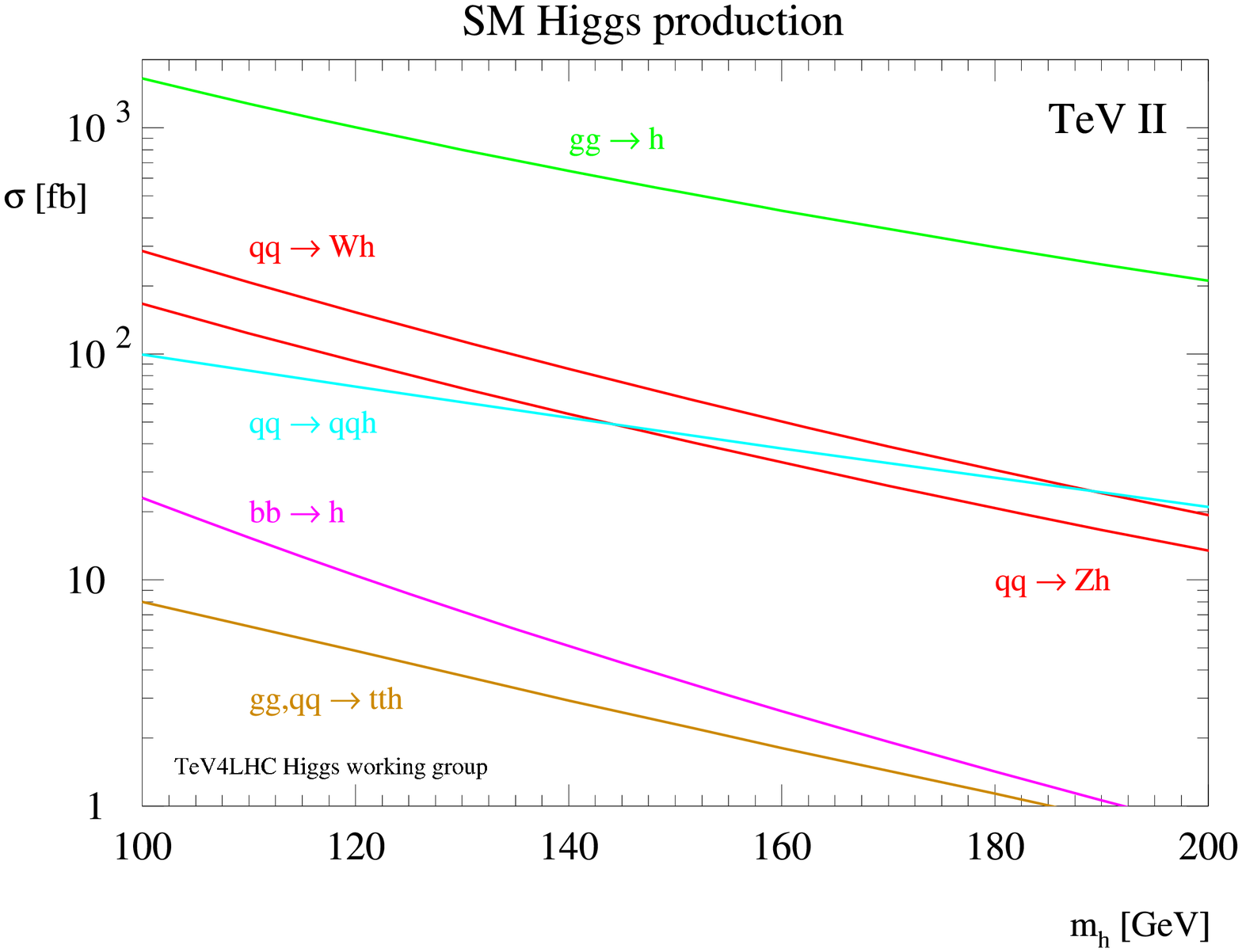,width=6.7cm}
\psfig{file=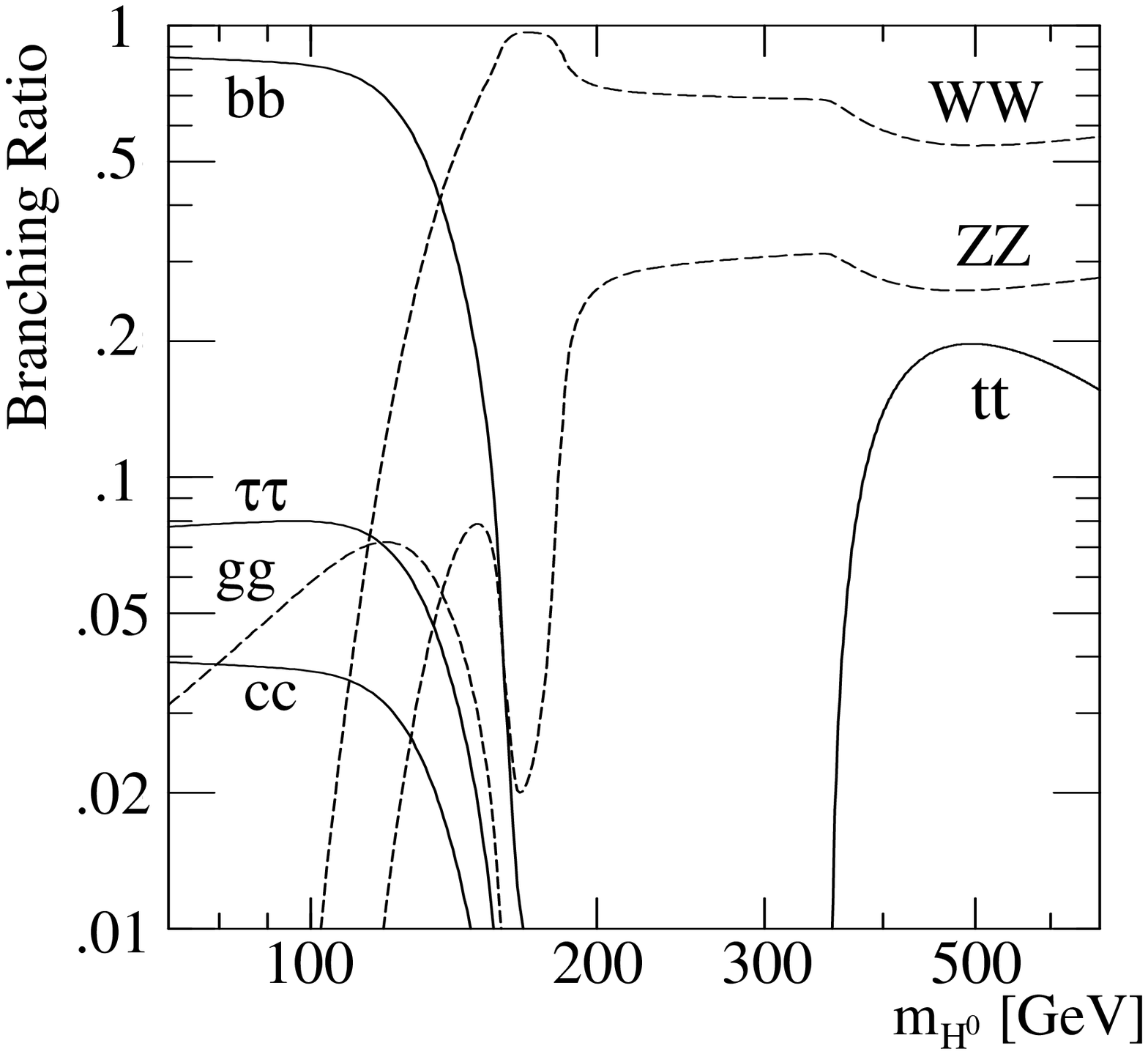,width=6.7cm}} 
\vspace*{8pt}
\caption{SM Higgs production cross section at the Tevatron (left) and its
decay branching ratio (right) as a function of the 
Higgs boson mass.\label{fig:xbr-decay}}
\end{figure}

The dominant Higgs production processes at the Tevatron are the 
gluon-gluon fusion
($gg\rightarrow H$) and the associated production with a $W$ or $Z$ 
boson~\cite{higgs-xsec}. The cross section for the production of SM Higgs 
and its decays are summarized in Fig.~\ref{fig:xbr-decay} as a function of the 
Higgs mass between 100-200 GeV/c$^{2}$. 
The cross section for $WH$ production is twice that 
of $ZH$ and is about a factor of 10 smaller than $gg\rightarrow H$. 
The Higgs boson decay branching fraction is 
dominated by $H\rightarrow b\bar b$ for the low-mass Higgs ($m_H < 135$ 
GeV/c$^2$) and by $H\rightarrow W^+W^-$ or $Z Z^*$ for the high-mass 
Higgs ($m_H>135$ GeV/c$^2$). A search for a low-mass Higgs boson in the 
$gg\rightarrow H\rightarrow b\bar b$ channel is extremely challenging because
the $b\bar b$ QCD production rate is many orders of magnitude larger than the 
Higgs boson production rate. Requiring the leptonic decay of the associated 
$W$ or $Z$ boson greatly improves the expected signal over background ratio in 
these channels. As a result, the Higgs associated production with
$H\rightarrow b\bar b$ is the most promising channel for the low-mass Higgs
boson searches. For the high-mass Higgs, $H\rightarrow
W^+W^-$ modes with leptonic decay provide the greatest 
sensitivity. The secondary channels of $H\rightarrow\gamma\gamma$,
$H\rightarrow \tau^+\tau^-$, and $t\bar t H$ are also considered at 
the Tevatron. Finally, all the channels have to be combined to achieve the best 
SM Higgs sensitivity.

\section{Search Strategies} 

The challenge for the standard model Higgs search at the Tevatron is that 
the Higgs signal is so tiny compared to that of 
other SM processes with the same final 
states. The search strategies employed by the CDF and D0 collaborations are 
quite similar and have been evolving constantly over time.  
We first maximize the signal acceptances by using efficient 
triggers, excellent lepton identifications, and powerful $b$-tagging, which 
can improve the signal to the background ratio up to the 1\% level. 
Then we use multivariate analysis (MVA) to exploit the kinematic 
differences between the signal and background, which can further enhance the 
signal to the background ratio up to the 10\% level in the high score regions. 
The same strategies have been used to help discover the single-top and 
diboson processes at the Tevatron, which provide a solid ground for 
how to isolate a small signal out of the huge background. 

For the low-mass $H\rightarrow b\bar b$ signatures we look for a 
$b\bar b$ mass resonance produced in association with a $W$ or $Z$ boson 
where $W$ decays into $l\nu$ or $Z$ decays into $l^+l^-$ or $\nu\bar \nu$. 
The $WH\rightarrow lvbb$ is the most sensitive channel that gives 
one high $P_T$ lepton, large \met, and two $b$-jets. Before $b$-tagging, the 
sample is predominated by the $W$ + light-flavor jets, which provides ideal
control data to test the background modeling.  

For the high-mass $H\rightarrow W^+W^-$ signatures, we look for the Higgs 
boson decaying into a $W^+W^-$ pair in the inclusive Higgs events that lead 
to many interesting final states. The most sensitive channel is both W bosons
decaying leptonically that gives an opposite-signed dilepton, large \met, and 
some jets from the initial state radiation or other production processes.  
Because of the missed neutrinos in the final state, 
the Higgs mass can not be reconstructed. We have to rely on the event kinematic
that distinguishes signal from background. For example, the $\Delta\phi$ of 
two leptons from the Higgs decay prefers a smaller $\Delta\phi$ than the 
background due to the fact that the Higgs is a scalar particle.
We can further improve the separation between signal and background by 
combining the $\Delta\phi$ with other 
kinematic variables in the event using a multivariate discriminant. 

\section{Recent Improvements}

\begin{figure}[htpb]
\centerline{\psfig{file=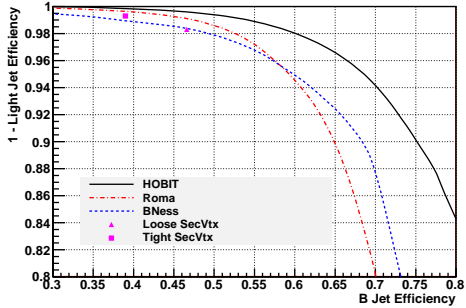,width=6.7cm}}
\vspace*{8pt}
\caption{A comparison of the purity-efficiency tradeoffs for HOBIT 
vs other $b$-taggers at CDF.\label{fig:beff-mistag}}
\end{figure}

Since there are two $b$-quark jets from the low-mass Higgs decay, improving 
$b$-tagging is crucial. Both CDF and D0 use MVA $b$-tagging
 to exploit the decay of 
long-lived $B$ hadron as displaced tracks/vertices. The typical efficiency is 
about 40-70\% with a mistag rate of 1-5\% per jet. 
Recently CDF combined their existing $b$-taggers into a Higgs Optimized  
$b$-tagger (HOBIT)~\cite{hobit} using a neural network tagging algorithm, 
based on sets of kinematic variables sensitive to displaced decay vertices and 
tracks within jets with large transverse impact
parameters relative to the hard-scatter vertices. 
Using an operating point 
which gives an equivalent rate of false tags, the new algorithm improves upon 
previous $b$-tagging efficiencies by $\approx$ 20\%. 
Fig.~\ref{fig:beff-mistag} shows the comparison of $b$-tag efficiency vs mistag
rejection for the existing taggers and HOBIT $b$-tagger as shown in the 
black curve. The $b$-tag efficiency is calibrated using the $t\bar t$ events 
selected in the $W$+ three or more jets sample 
and the $b$-enriched inclusive electron data, while the mistag rate is 
determined using the $W$ + one jet 
sample. The ratio of $b$-tag efficiency per $b$-jet measured 
from the data and the Monte Carlo is used as a scale factor to correct for 
the differences in the Monte Carlo modeling, 

To discriminate Higgs signal events against background, using MVA 
would improve the background rejection 
with a sensitivity gain of 25\%, compared to using a single variable alone, 
such as the dijet mass. We can further improve MVA by training 
different backgrounds, splitting events into sub-channels based 
on S/B, e.g. lepton type, number of jets. 
CDF trains $ ZH\rightarrow llbb$ separately against $t\bar t$, $Z+c\bar c$ or 
$c$, and diboson to build the final discriminat.
 

\section{Low-Mass Searches}
  We describe the searches for the low-mass Higgs boson at the Tevatron 
  in some detail.   
\subsection{$WH\rightarrow l\nu b\bar b$} 

One of the golden channels for the low-mass Higgs boson search is the Higgs 
produced in association with a $W$ boson with 
$WH\rightarrow l\nu b\bar b$~\cite{cdfwh, d0wh}. We select events with one 
isolated high $P_T$ lepton (electron, muon, or isolated track), a large 
missing transverse energy, and two or three jets, of which at least one is 
required to be $b$-tagged as containing a weakly-decaying B hadron. 
Events with more than one isolated lepton are rejected. For the multivariate 
discriminant, CDF trained a Bayesian neural network discriminant(BNN) in 
the $W$ + two and three jets for each Higgs mass, separately for each lepton, 
jet multiplicity, $b$-tagging category. 
For the D0 $WH\rightarrow l\nu b\bar b$ analyses, 
the data are split by lepton type, jet multiplicity, and the number of 
$b$-tagged jets, similar to CDF. The outputs of boosted decision trees (BDT), 
trained separately for each sample and for each Higgs boson mass, are used as 
the final discriminating variables. 

\begin{figure}[htpb]
\centerline{\psfig{file=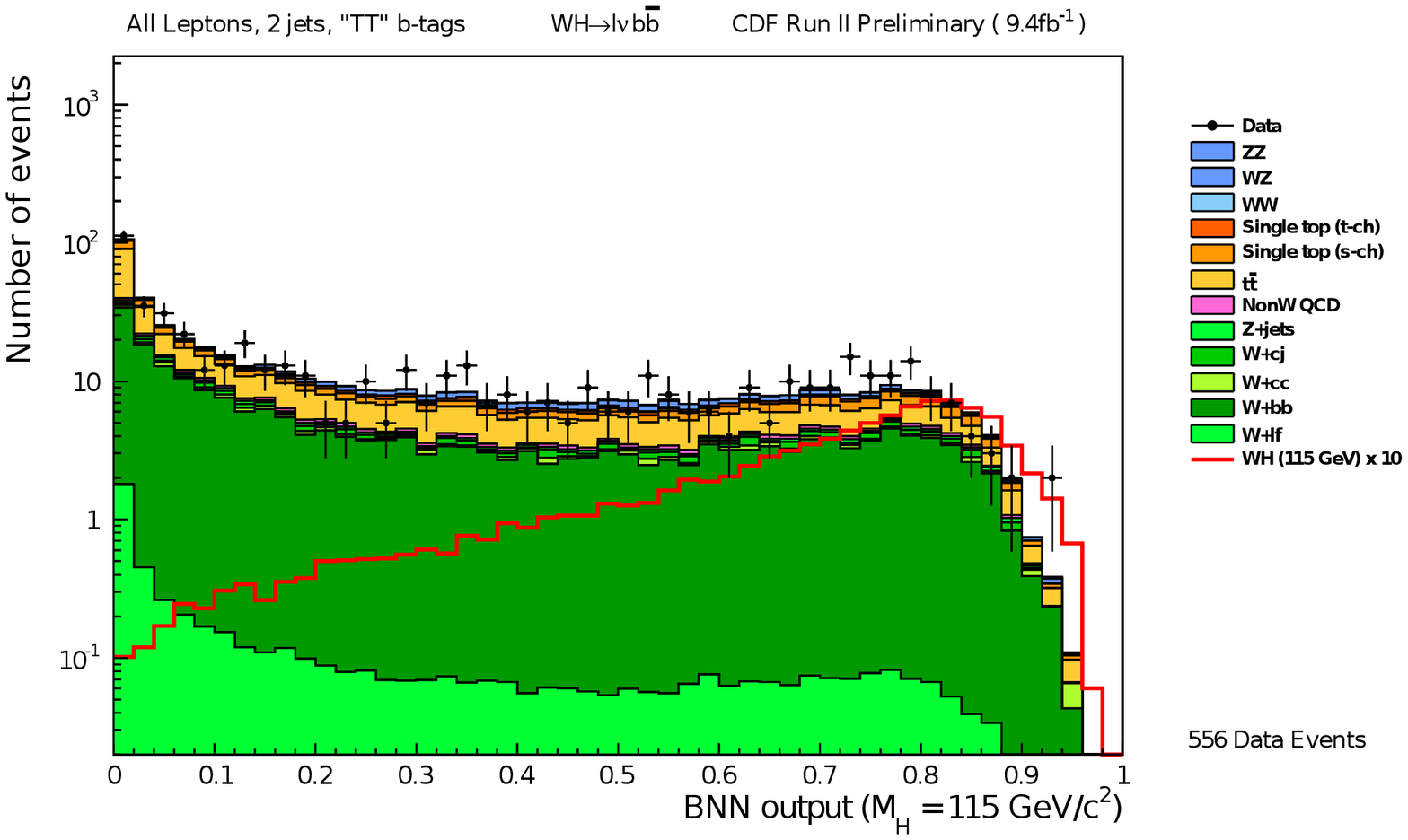,width=6.7cm}
\psfig{file=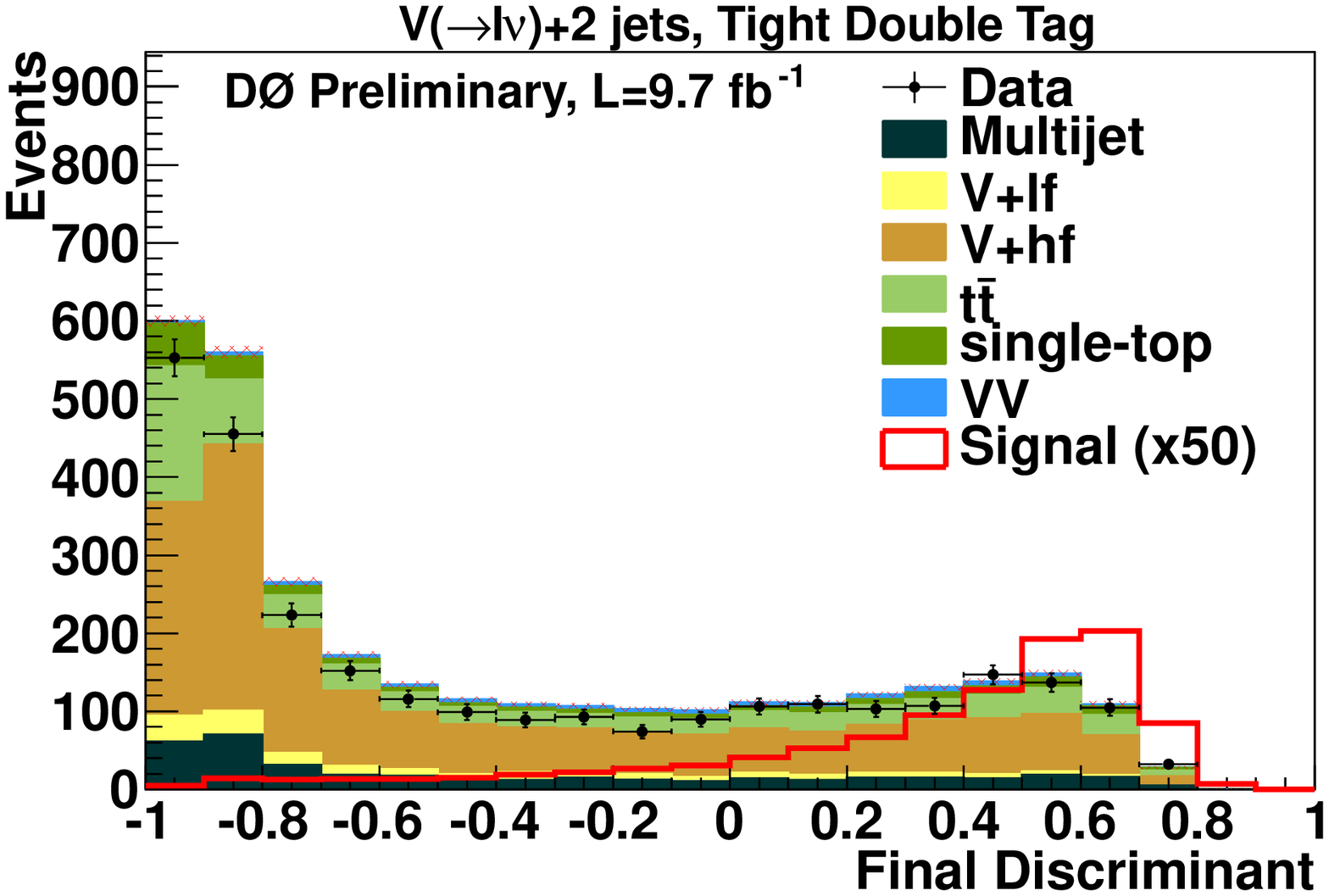,width=6.7cm}}
\vspace*{8pt}
\caption{The final discriminants for a 115 GeV/c$^2$ Higgs signal 
are shown in the $W$ + two jets 
after 2 $b$-tags for CDF's BNN (left) and D0's BDT (right), 
respectively.\label{fig:wh-output}}
\end{figure}

We perform a direct search for an excess of events in the signal
region of the final discriminant from each event category.
Fig.~\ref{fig:wh-output} shows the output of the final discriminants 
optimized for a 115 GeV/c$^2$ Higgs signal in the 
double $b$-tagged $W$ + two jets data from CDF and D0, respectively. 
The data and background predictions are in good agreement. The expected Higgs
signals are also shown, but rescaled by a large factor. 

\begin{figure}[htpb]
\centerline{\psfig{file=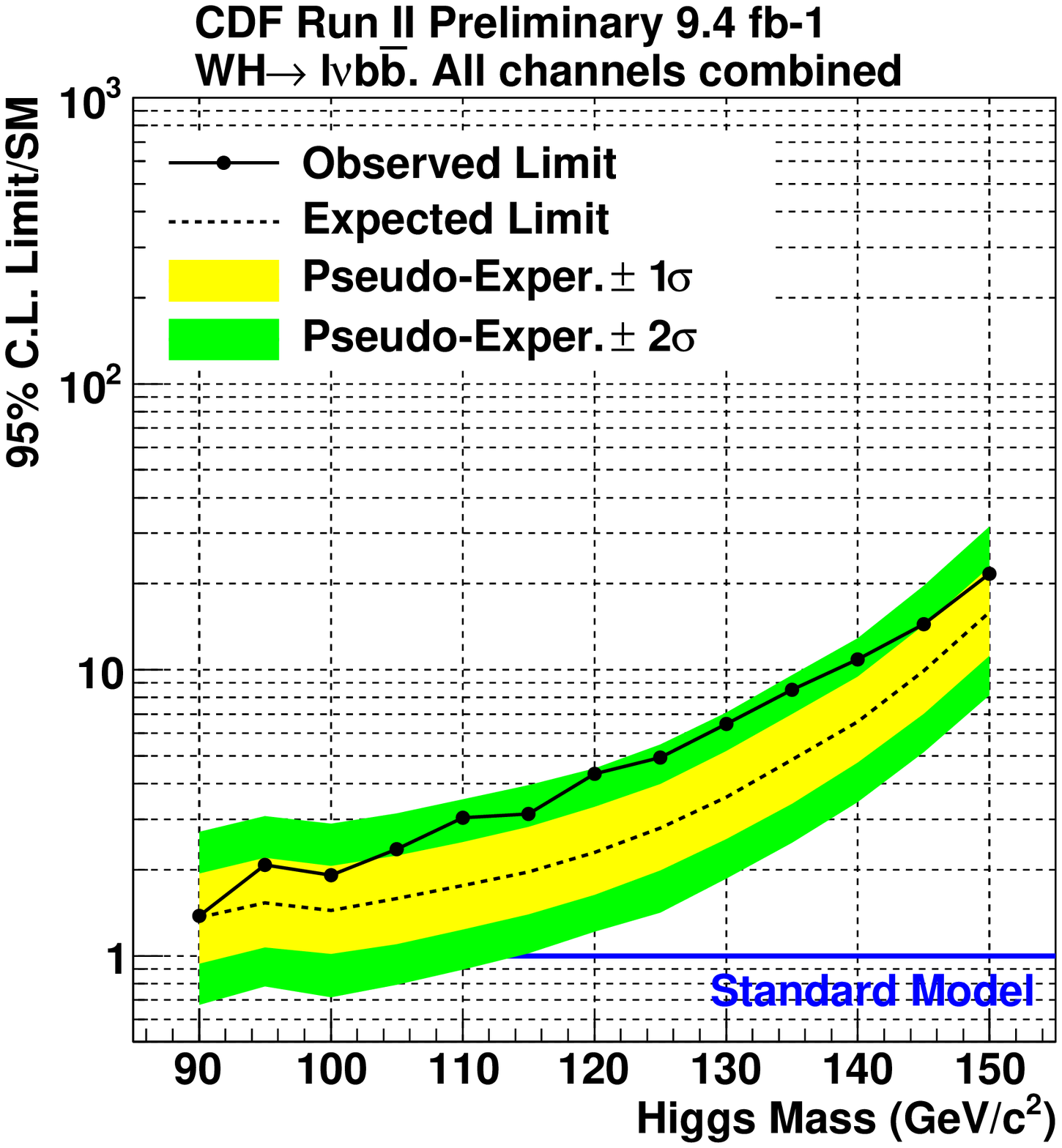,width=6.7cm}
\psfig{file=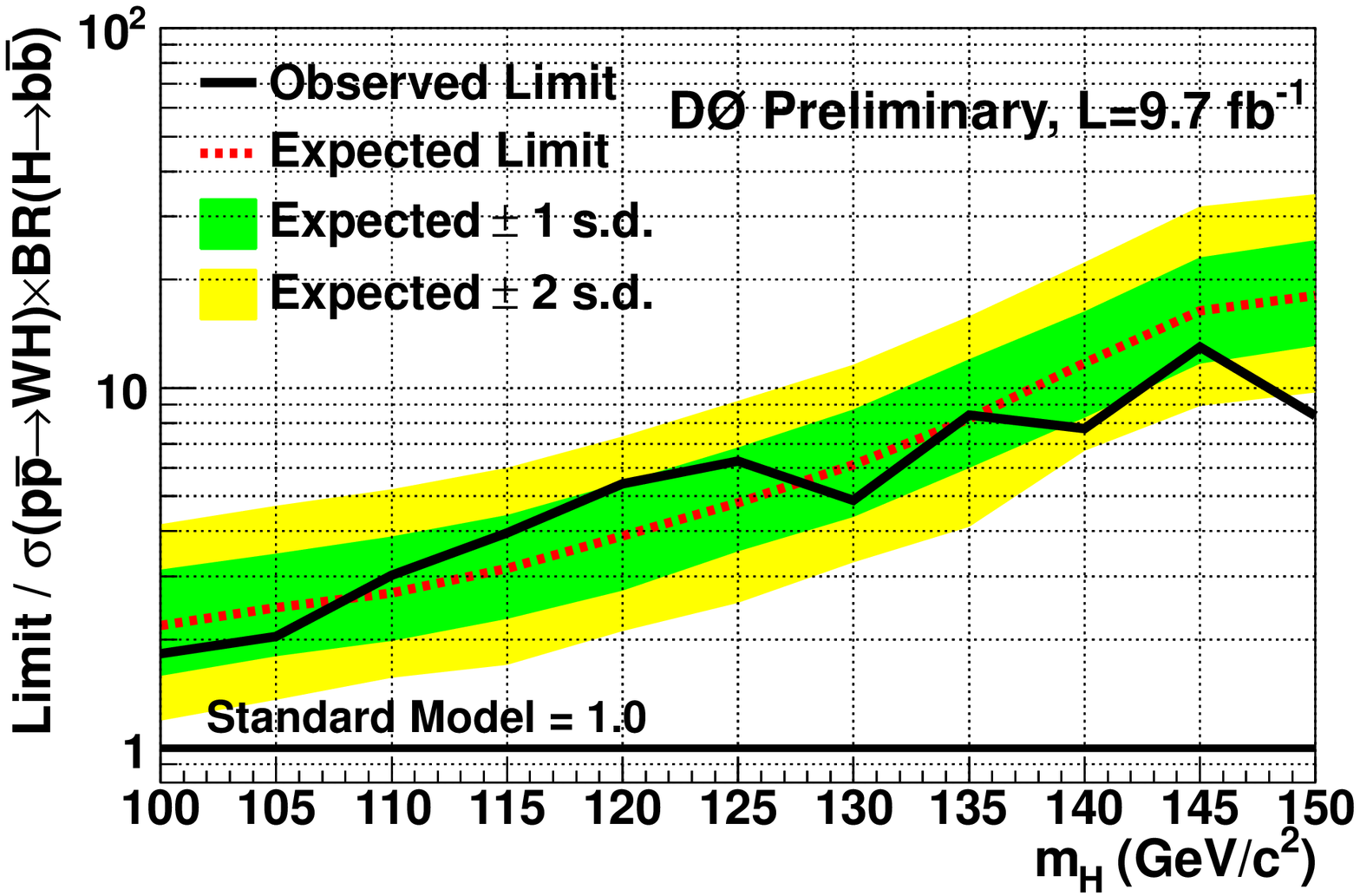,width=6.7cm}}
\vspace*{8pt}
\caption{Observed and expected 95\% CL upper limits on SM Higgs production
as a function of Higgs boson mass in the $WH\rightarrow l\nu b\bar b$ from 
CDF (left) and D0 (right), respectively.\label{fig:whlimits}}
\end{figure}

Since there is no significant excess of signal observed in the data, we set 
an upper limit at 95\% CL on the Higgs production cross section times 
branching ratio with respect to the SM predictions as a function of Higgs 
mass, as shown in Fig.~\ref{fig:whlimits}.
For $m_H=125$ GeV/c$^2$, CDF set an observed (expected) upper limit at 4.9(2.8) 
while D0 set a limit at 6.2(4.8). They are not yet competitive for a single 
channel and we need to combine all other channels including both CDF and D0 
results together. 

\subsection{$ZH\rightarrow l^+l^- b\bar b$}

\begin{figure}[htpb]
\centerline{\psfig{file=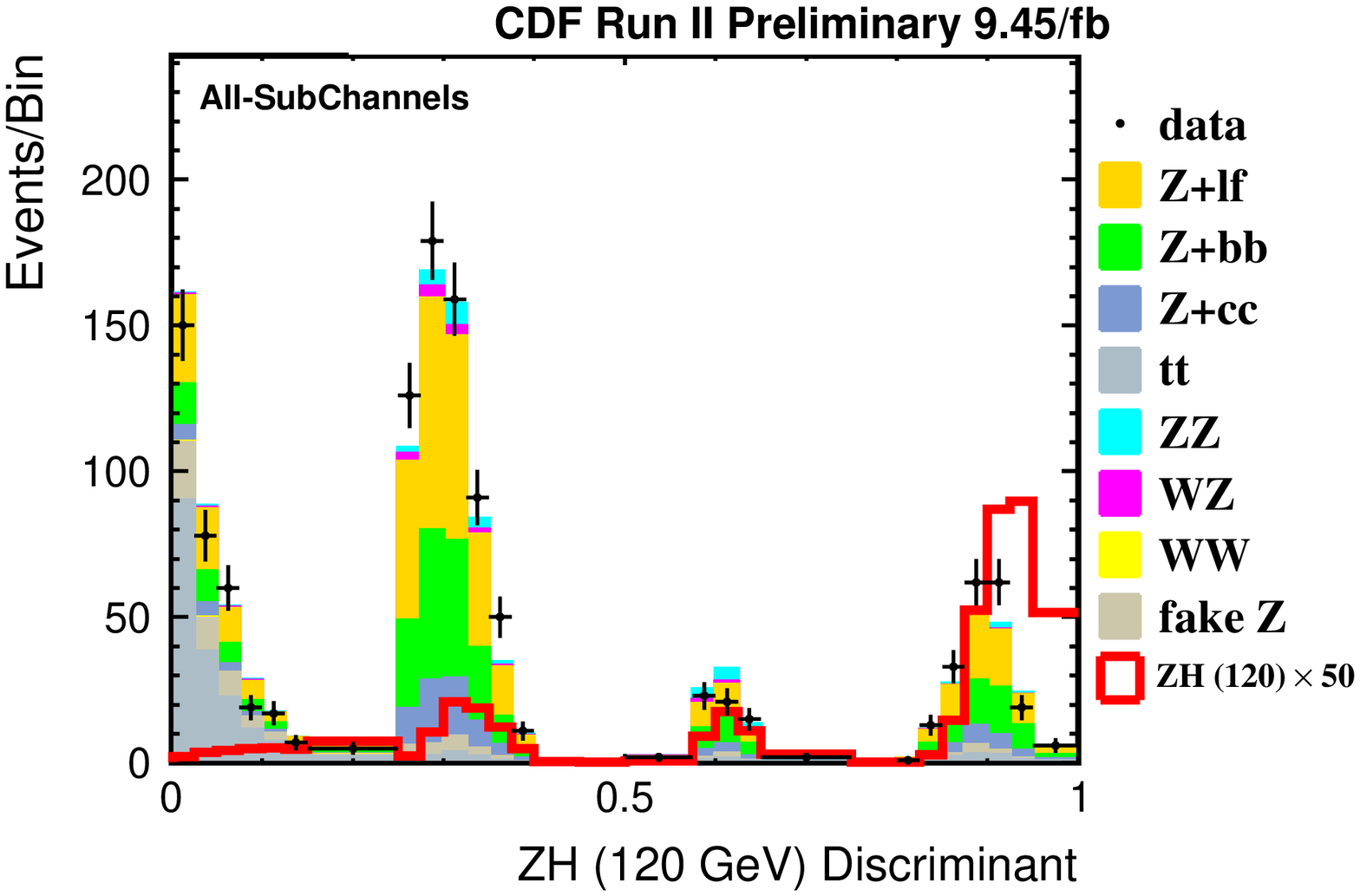,width=6.7cm}
\psfig{file=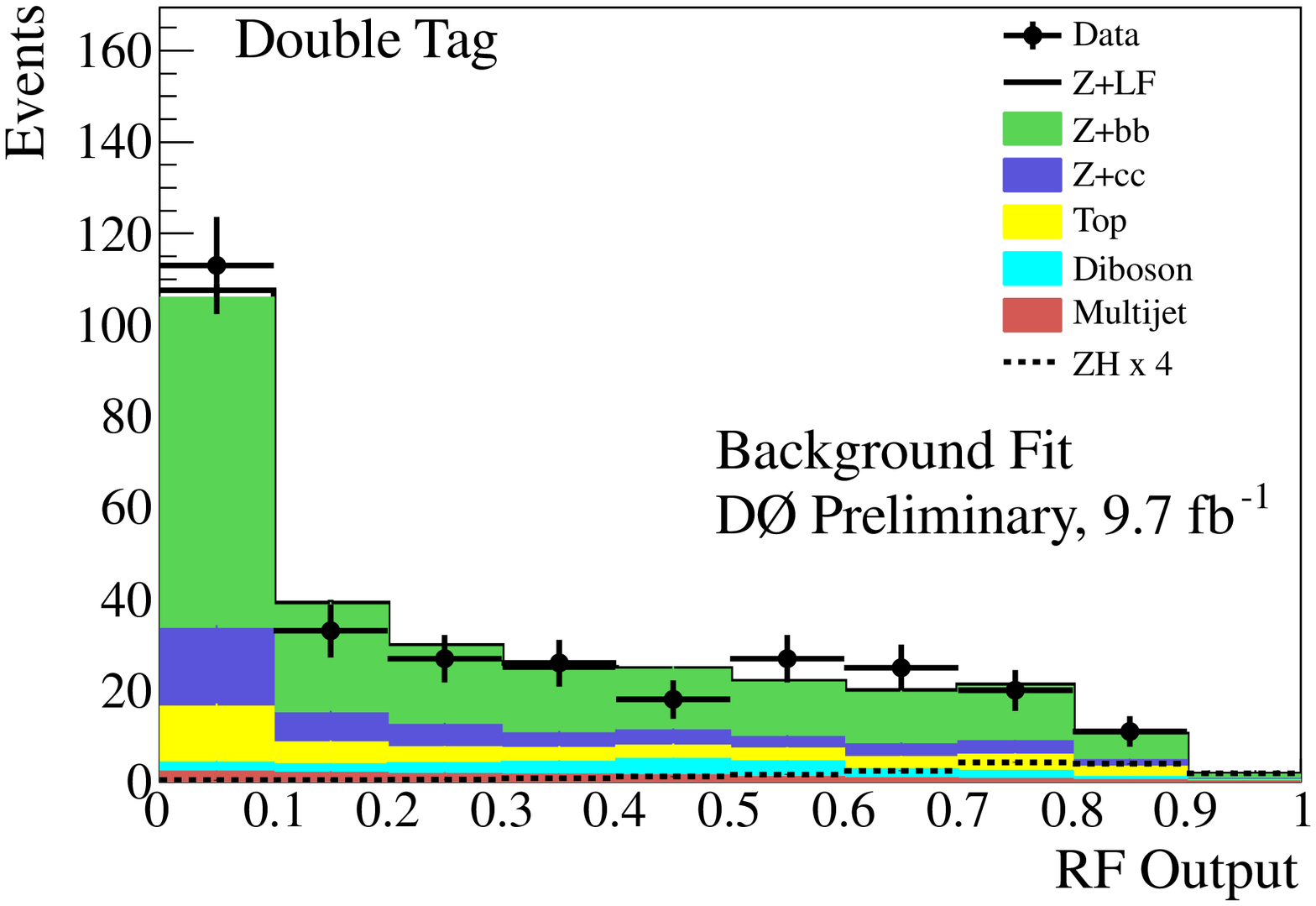,width=6.7cm}}
\vspace*{8pt}
\caption{The final discriminants are 
shown in the $b$-tagged $Z$ + two or three jets events from  
CDF (NN output, left) and the double $b$-tagged $Z$ + two jets events from D0 
(RF output, right), respectively.\label{fig:llbb-output}}
\end{figure}

Another interesting channel to search for the low-mass Higgs boson is 
$ZH\rightarrow l^+l^-b\bar b$~\cite{cdfllbb,d0llbb}. It provides a clean
signature, but has a low event 
yield due to a small branching fraction of 
$Z\rightarrow e^+e^-$ and $\mu^+\mu^-$.
We select events with two high $P_T$ leptons from $Z$ decay and
two or three jets. Events are further divided based on lepton type, 
jet multiplicity, and the number of $b$-tagged jets, similar to 
$WH\rightarrow l\nu b\bar b$. To increase signal accepance D0 loosens the 
selection criteria for one of the leptons to include an isolated track not 
reconstructed in the muon detector or an electron from the inter-cryostat 
region of the D0 calorimeter. 
CDF uses neural networks to select loose dielectron
and dimuon candidates. D0 applies a kinematic fit to optimize reconstruction 
while CDF corrects jet energies for the missing $E_T$ using a neural network
approach. D0 uses random forests of decision trees to provide the final 
discriminant for sitting limits. CDF utilizes a multi-layer discriminant based
on neural networks where separate discriminant functions are used to define
four separate regions of the final discriminant function. 
Fig.~\ref{fig:llbb-output} shows the final discriminant optimized for 
a Higgs signal ($m_H=115$ GeV/c$^2$) in the $b$-tagged events from CDF and 
the double $b$-tagged events from D0, respectively. 
There seem to be some excess of events in the high score signal 
region, but not statistically significant yet. CDF set an observed (expected) 
upper limit at 95\% CL on the Higgs cross section times branching ratio over 
the standard model prediction at 7.2(3.6) while D0 set a limit at 6.9(5.9) 
for the Higgs mass at 125 GeV/c$^2$. 

\subsection{$WH, ZH\rightarrow $\met$b\bar b$}

\begin{figure}[htpb]
\centerline{\psfig{file=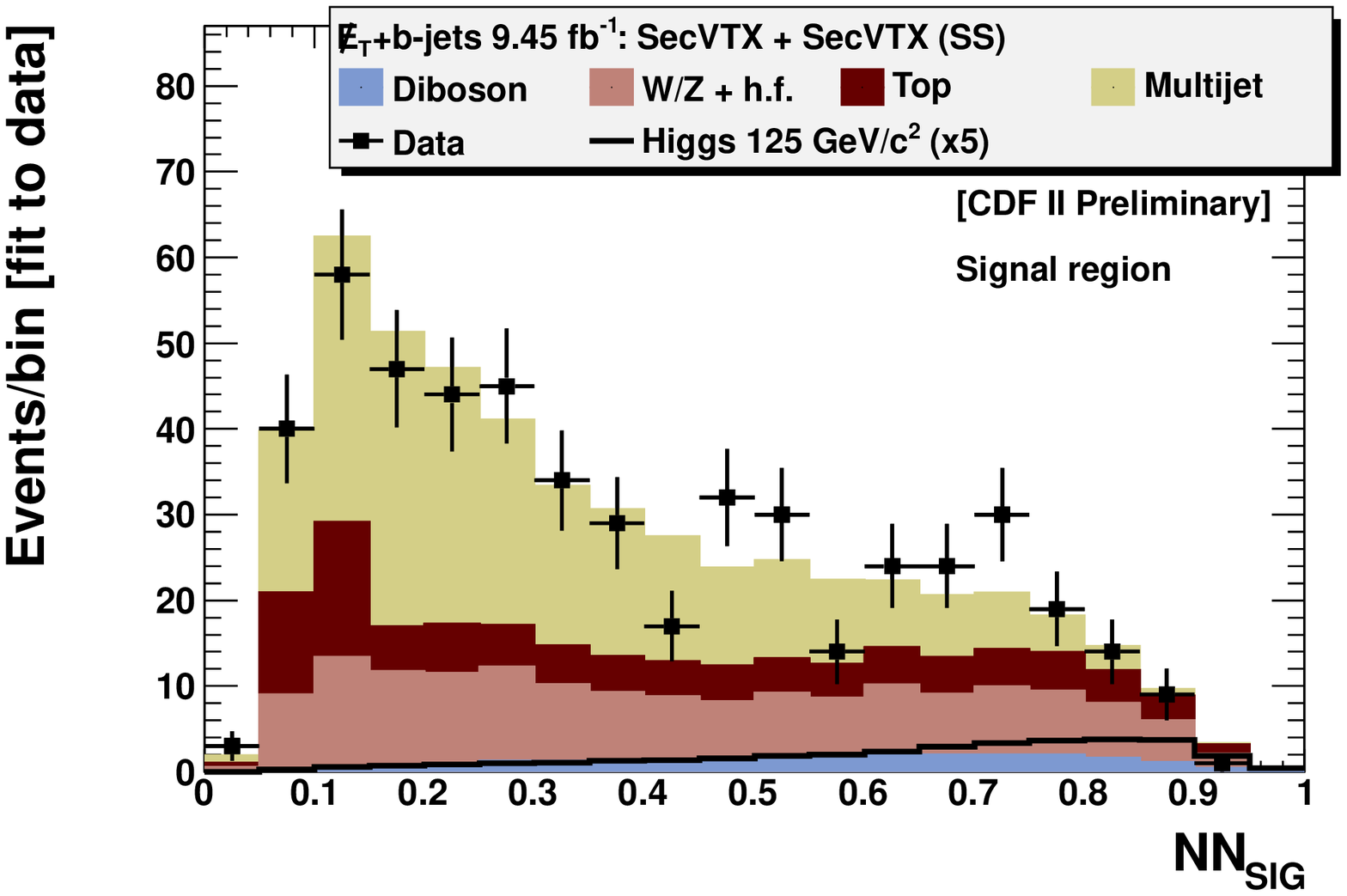,width=6.7cm}
\psfig{file=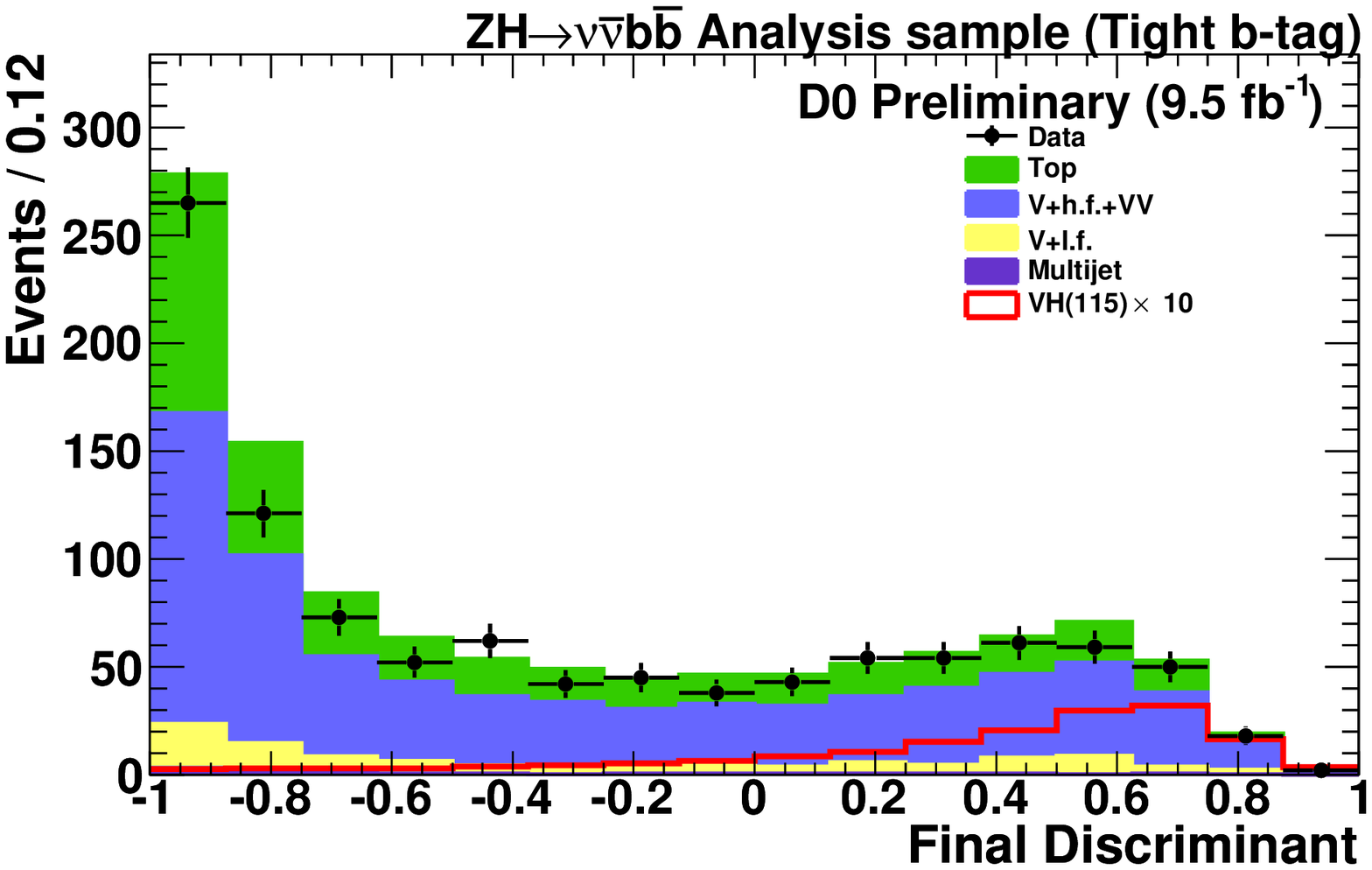,width=6.7cm}}
\vspace*{8pt}
\caption{The final discriminants are shown in the \met + two jets after 2 
$b$-tags for CDF (left) and after tight tag for D0 (right), 
respectively.\label{fig:metbb-output}}
\end{figure}

We also looked for the Higgs boson in the $ZH$ and $WH$ channels where 
the $Z$ decays into two neutrinos or the lepton from $W$ decay is 
undetected~\cite{cdfmetbb,d0metbb}. It has a large signal rate as well as a 
large QCD-multijet background. However, the final state is relatively clean, 
containing two high $E_T$ jets and a large missing transverse energy. We require
\met$> 50 $ GeV and two $b$-tagged jets. Both CDF and D0 use a track-based 
missing transverse momentum calculation as a discriminant against false \met.
In addition both CDF and D0 utilize multivariate technique, a neural network 
for CDF and a boosted decision tree for D0, to further discriminate against the 
multi-jet background.  The final discriminant is obtained 
for a Higgs signal ($m_H= 115$ GeV/c$^2$) by combining dijet 
mass, track \met, and other kinematic variables, which are shown in 
Fig.~\ref{fig:metbb-output} for CDF and D0, respectively.
There seems a good agreement between data and background predictions,
as CDF set an 
observed (expected) upper
limit at 95\% CL on the Higgs cross section times branching ratio over the 
standard model prediction at 6.8(3.6) while D0 set a limit at 3.8(4.3) 
for the Higgs mass at 125 GeV/c$^2$. 

\section{High-Mass Searches} 
\begin{figure}[htpb]
\centerline{\psfig{file=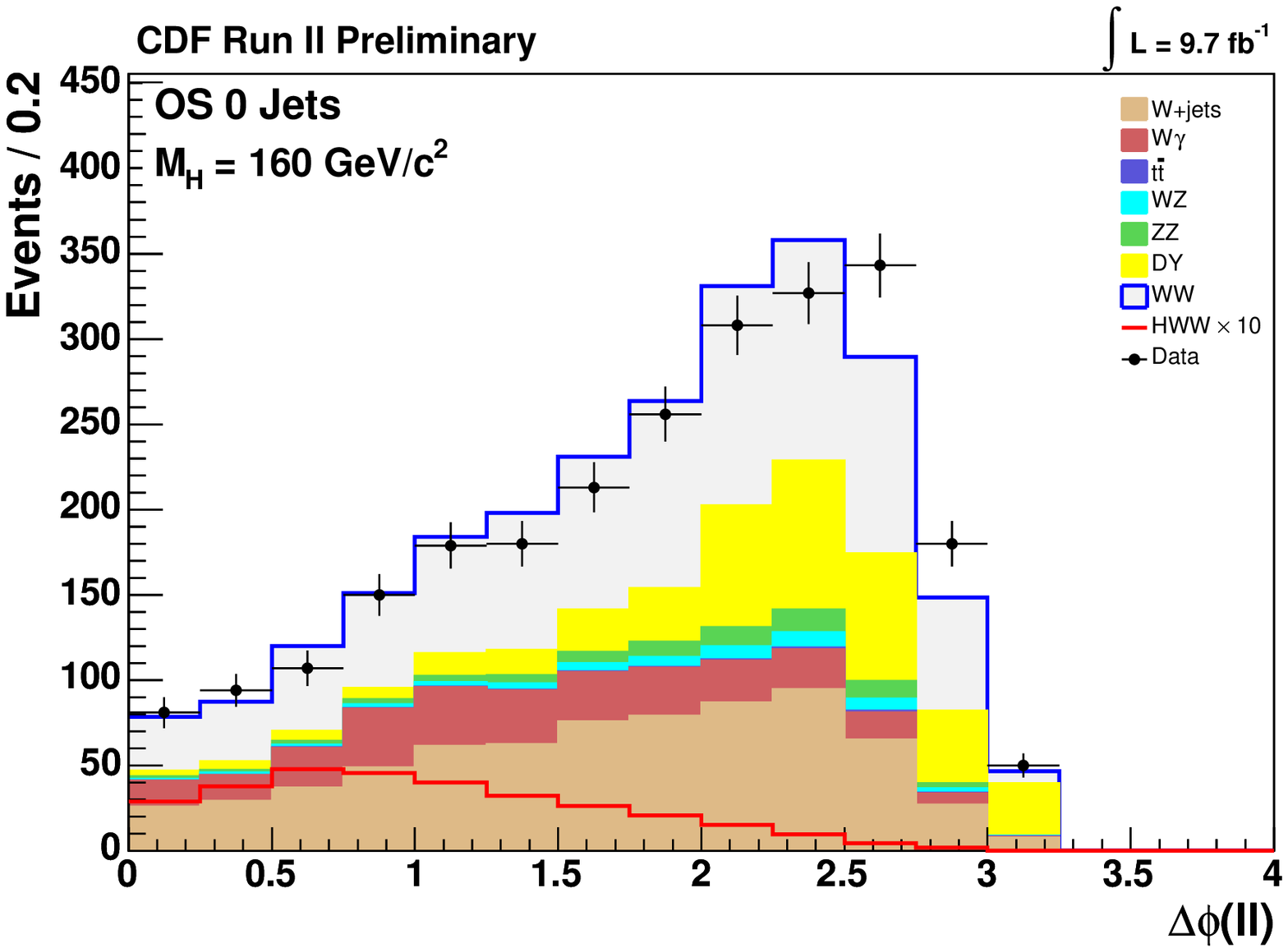,width=6.7cm}
\psfig{file=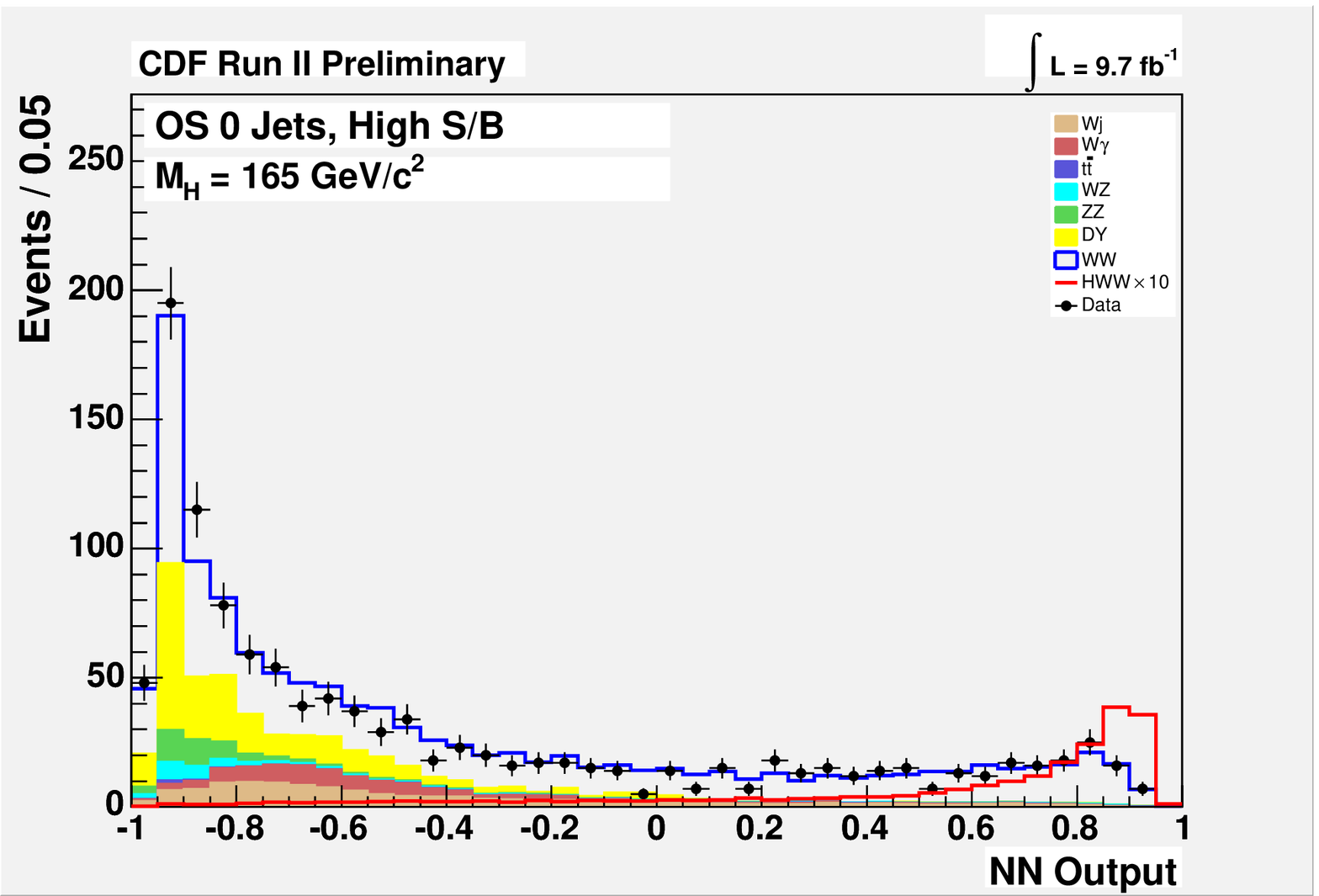,width=6.7cm}}
\vspace*{8pt}
\caption{The $\Delta\phi$ distribution of the two opposite-signed leptons in 
the events with no jets (left) and the final multivariate discriminant 
in right.\label{fig:dphicdf}}
\end{figure}

\begin{figure}[htpb]
\centerline{\psfig{file=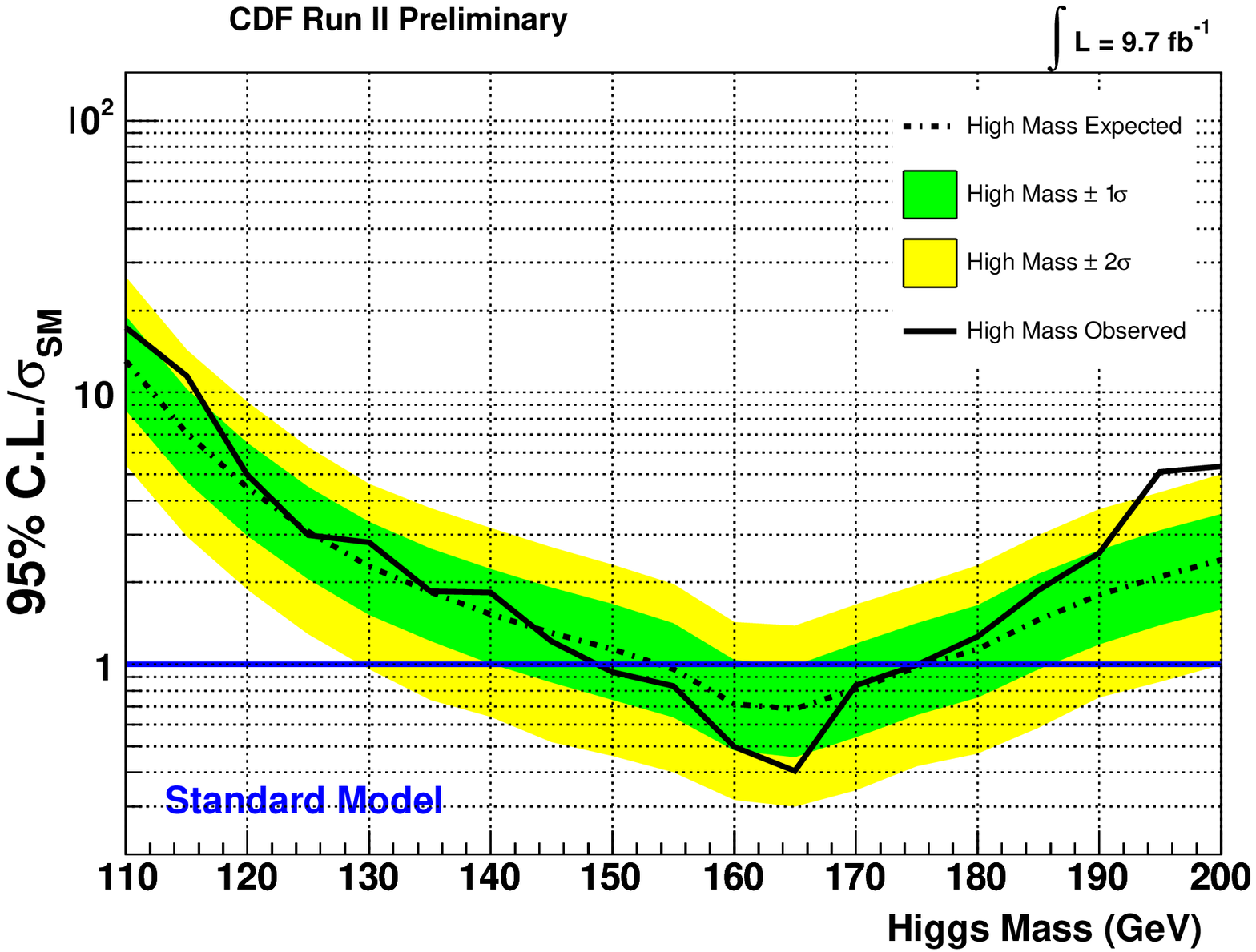,width=6.7cm}
\psfig{file=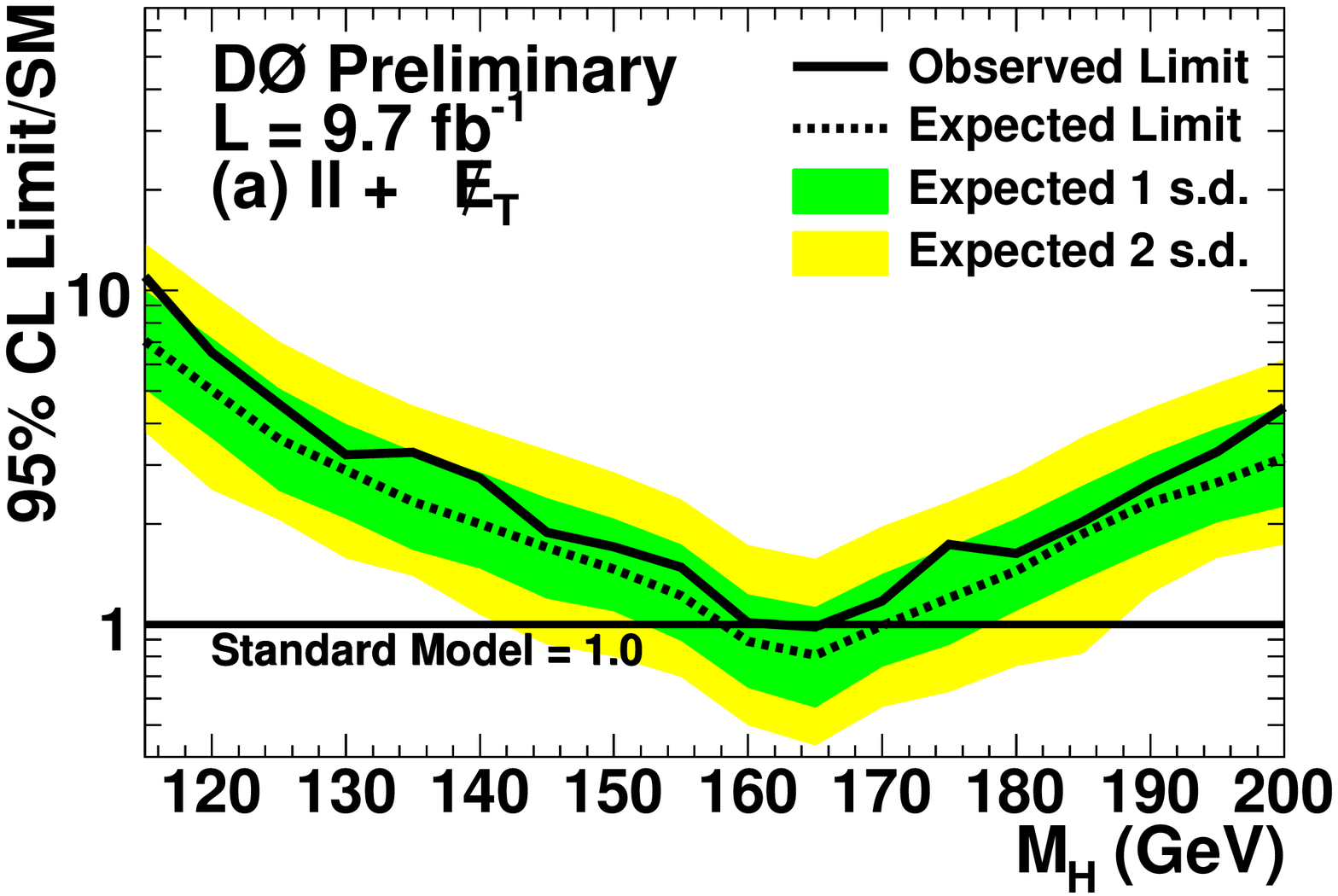,width=6.7cm}}
\vspace*{8pt}
\caption{Observed and expected 95\% CL upper limits on SM Higgs production
as a function of Higgs boson mass in the $H\rightarrow W^+W^-$ from 
CDF (left) and D0 (right), respectively.\label{fig:wwlimit}}
\end{figure}

For the high-mass  signatures, 
we look for Higgs decay into WW pair in the 
inclusive Higgs events that lead to many interesting final 
states~\cite{cdfww,d0ww}.
The most sensitive channel is both W decaying leptonic that gives an  
opposite-signed dilepton pair, large missing Et, and some jets from the initial
state radiation. The presence of neutrinos in the final state prevents the 
precise reconstruction of the Higgs boson mass. We have to rely on the event 
kinematic that distinguishes signal from background based on 
the scalar nature of 
the Higgs boson. We also include the processes $WH\rightarrow WW^+W^-$ 
and $ZH\rightarrow Z W^+W^-$ that 
give rise to like-sign dilepton and trilepton in the final states. 
Fig.~\ref{fig:dphicdf} shows the $\Delta\phi$ distribution of 
two opposite-signed leptons in the zero jet bin. 
The red line is for the signal, 
which prefers a smaller 
$\Delta\phi$ than most backgrounds. By combining the $\Delta\phi$ with other 
kinematics we obtain a multivariate discriminant 
shown in Fig.~\ref{fig:dphicdf} 
that improves the analysis significantly. 
We set a 95\% upper limit on the production cross section times branching 
ratio over the standard model prediction as a function of the tested Higgs 
mass after combing all $H\rightarrow W^+W^-$ channels including the low mass 
dileptons, the same sign, and trileptons from $WH$ and $ZH$, as shown in 
Fig.~\ref{fig:wwlimit}. CDF observes some deficit near 165 GeV/c$^2$ while 
D0 observes a broad excess, but they are consistent with each other. 

\section{Secondary Searches} 

\begin{figure}[htpb]
\centerline{\psfig{file=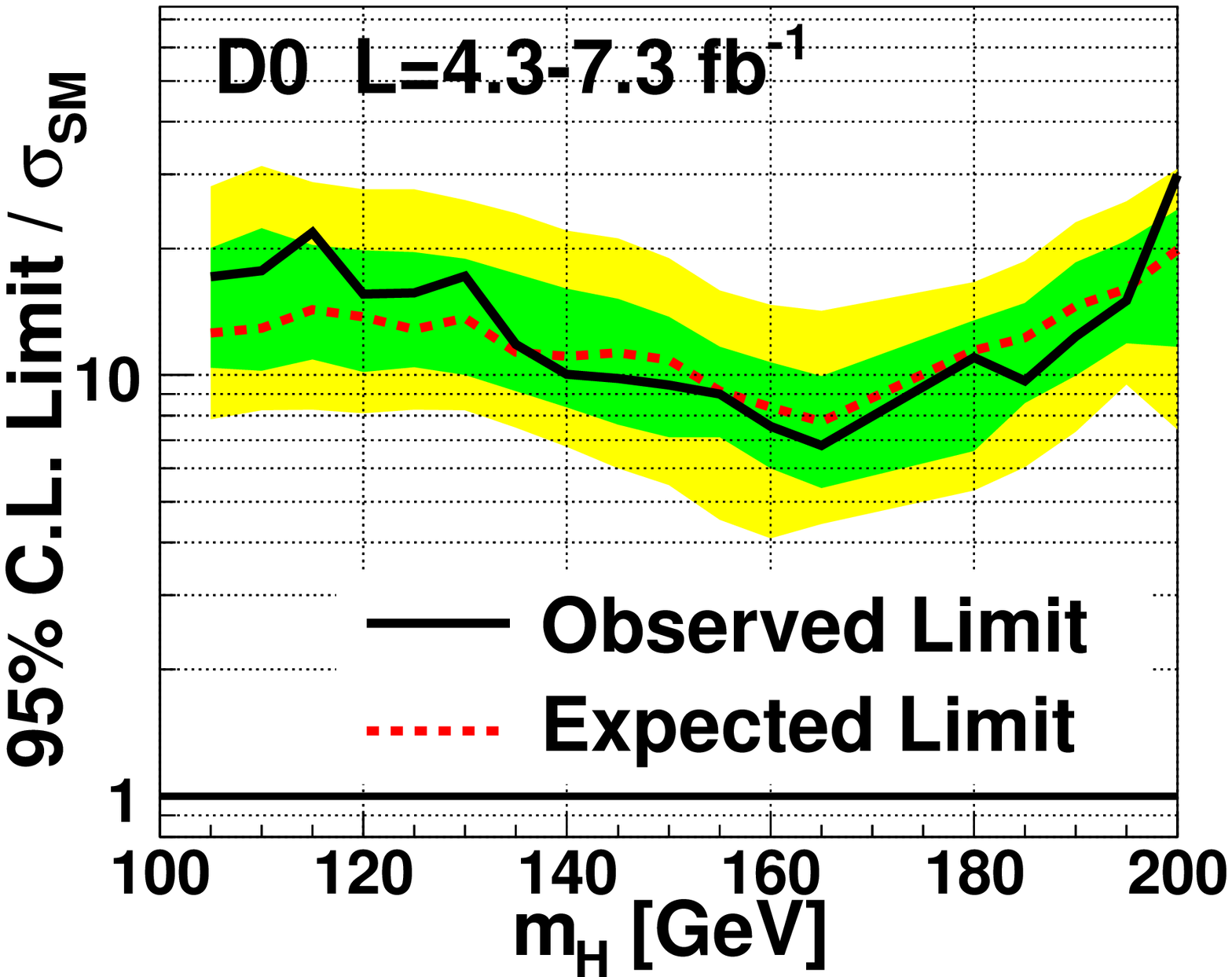,width=4.7cm}
\psfig{file=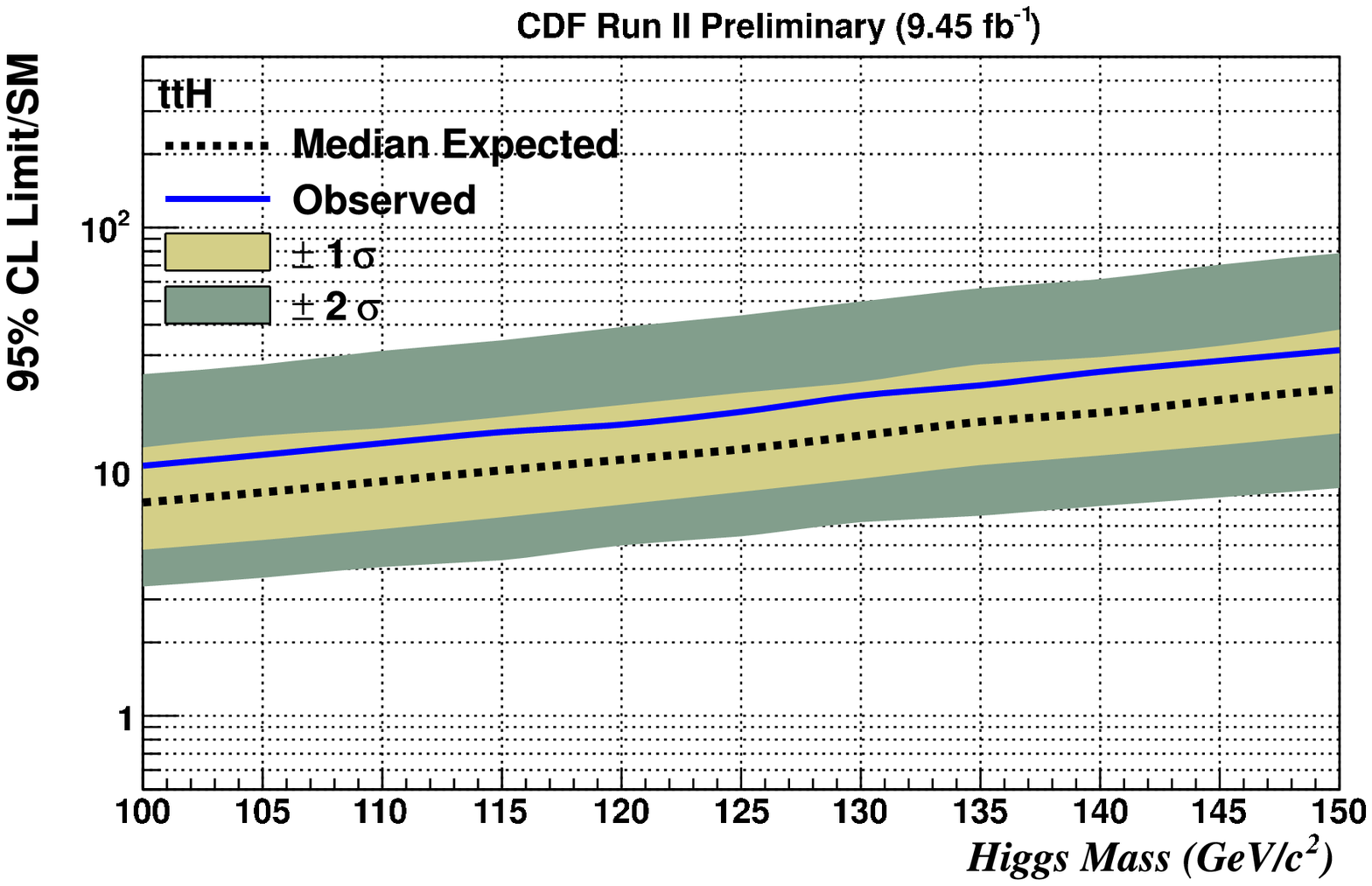,width=4.7cm}
\psfig{file=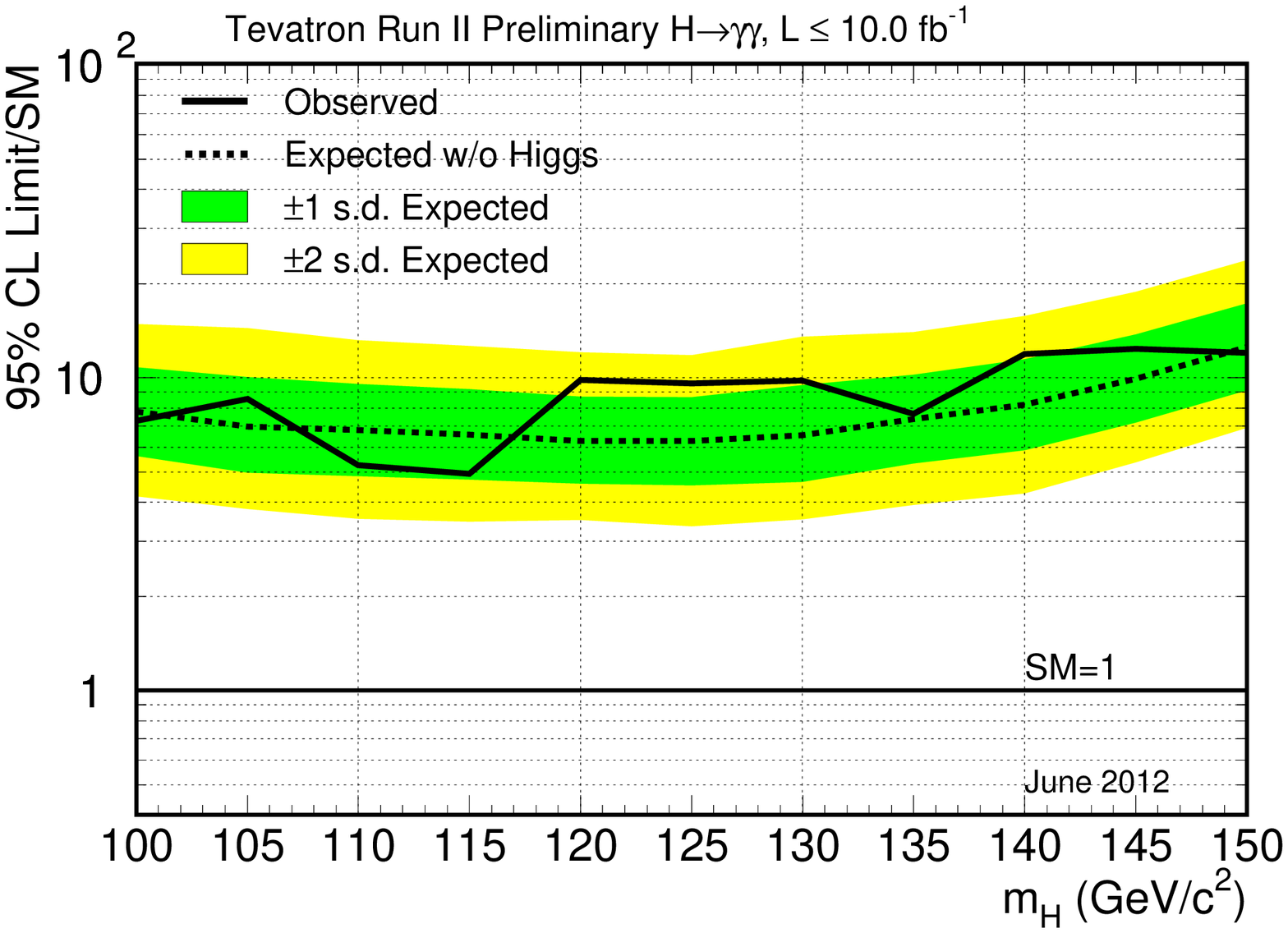,width=4.7cm}}
\vspace*{8pt}
\caption{The limits obtained from the searches on $H\rightarrow \tau^+\tau^-$ 
from D0 (left), $t\bar t H$ from CDF (middle), and $H\rightarrow \gamma\gamma$
from the Tevatron (left), respectively.\label{fig:secondsearch}}
\end{figure}

Other searches are also considered for the 
$H\rightarrow \tau^+\tau^-$ decay~\cite{cdftau,d0tau}, the 
$H\rightarrow\gamma\gamma$ decay~\cite{cdfgg,d0gg}, and the $t\bar t H$
production~\cite{cdftth}. Fig.~\ref{fig:secondsearch}
shows their 95\% CL upper limits on the production cross section times 
branching ratio with respect to the SM prediction, which is about 
a factor of ten larger than the SM Higgs sensitivity. But they do help 
to achieve the best Higgs sensitivity at the Tevatron.  

\section{Tevatron Combinations} 
We have searched for all possible SM Higgs production and decays. and set 
limits with respect to nominal SM predictions.
CDF and D0 are in good agreement and we combine them to improve 
the Tevatron final Higgs sensitivity.

\begin{figure}[htpb]
\centerline{\psfig{file=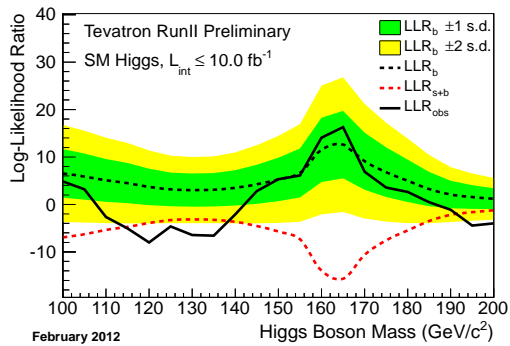,width=6.7cm}}
\vspace*{8pt}
\caption{The Tevatron combined LLR distributions as a function of 
Higgs mass.\label{fig:LLR}}
\end{figure}

To check the Tevatron Higgs sensitivity, we use
the log-likelihood ratios (LLR) with different signal hypotheses to test the
expected sensitivity as a function of Higgs mass, as shown in Fig~\ref{fig:LLR}.
The black dot is for the background-only hypothesis, the red dot is for 
signal-plus-background hypothesis, and the solid curve is for the 
observed data. The colored bands
indicate 1 or 2 sigma width of LLR for the background only distribution. 
The separation between the background only and the signal + background 
provides a measure of the search sensitivity, which is about 2 sigma for 
the Higgs boson mass at 125 GeV/c$^2$. The data seem consistent with the 
signal + background hypothesis between 115 and 135 GeV/c$^2$.

\begin{figure}[htpb]
\centerline{\psfig{file=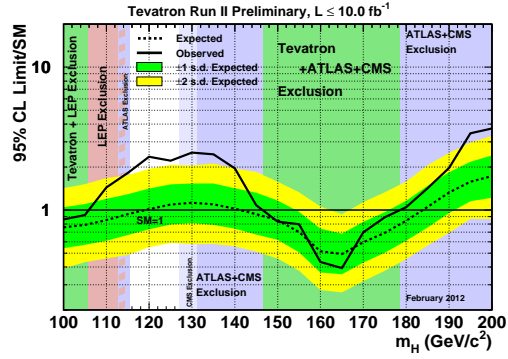,width=6.7cm}}
\vspace*{8pt}
\caption{The Tevatron combined Higgs limit as a function of 
tested Higgs mass.\label{fig:tevlimit}}
\end{figure}

\begin{figure}[htpb]
\centerline{\psfig{file=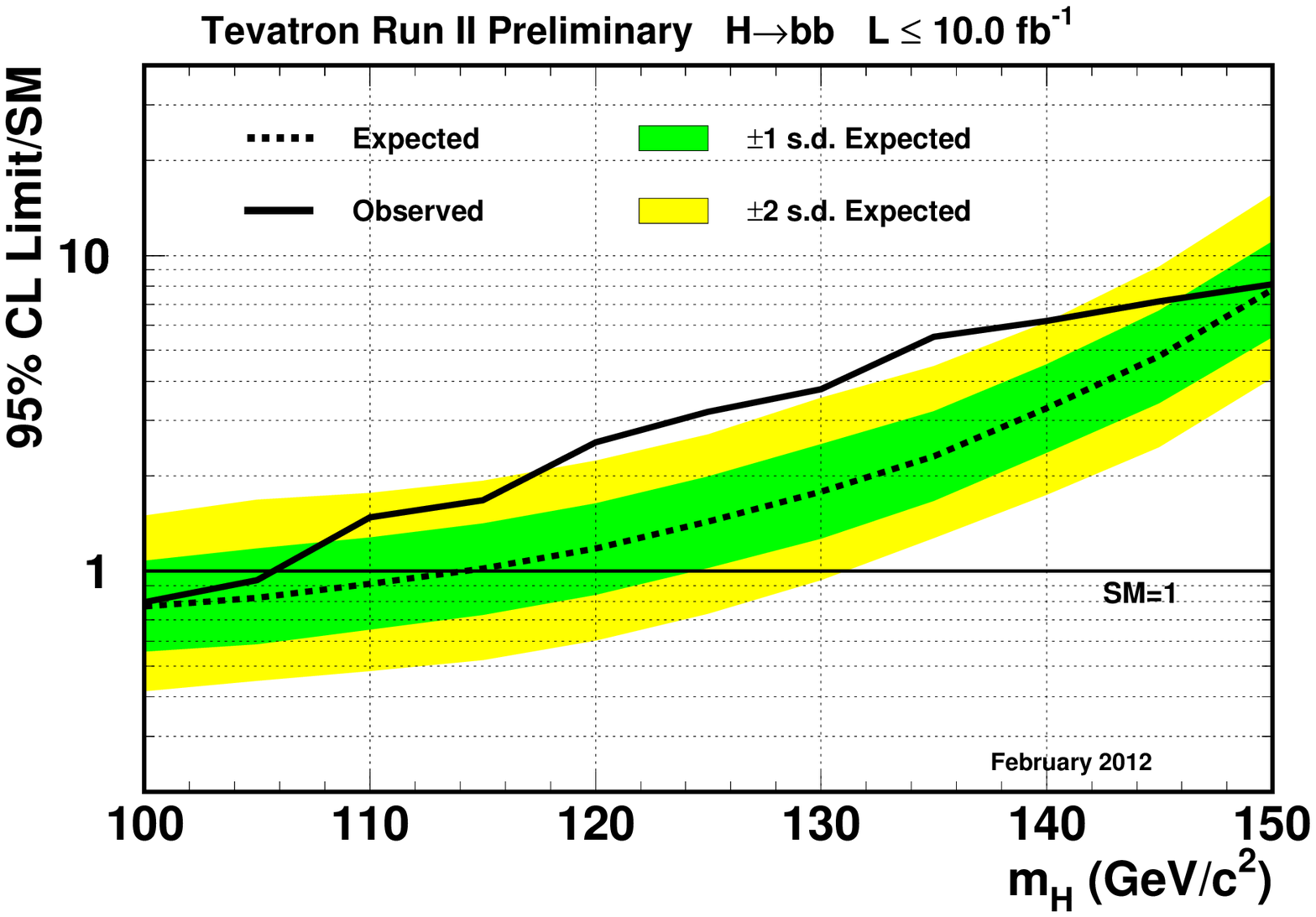,width=6.7cm}
\psfig{file=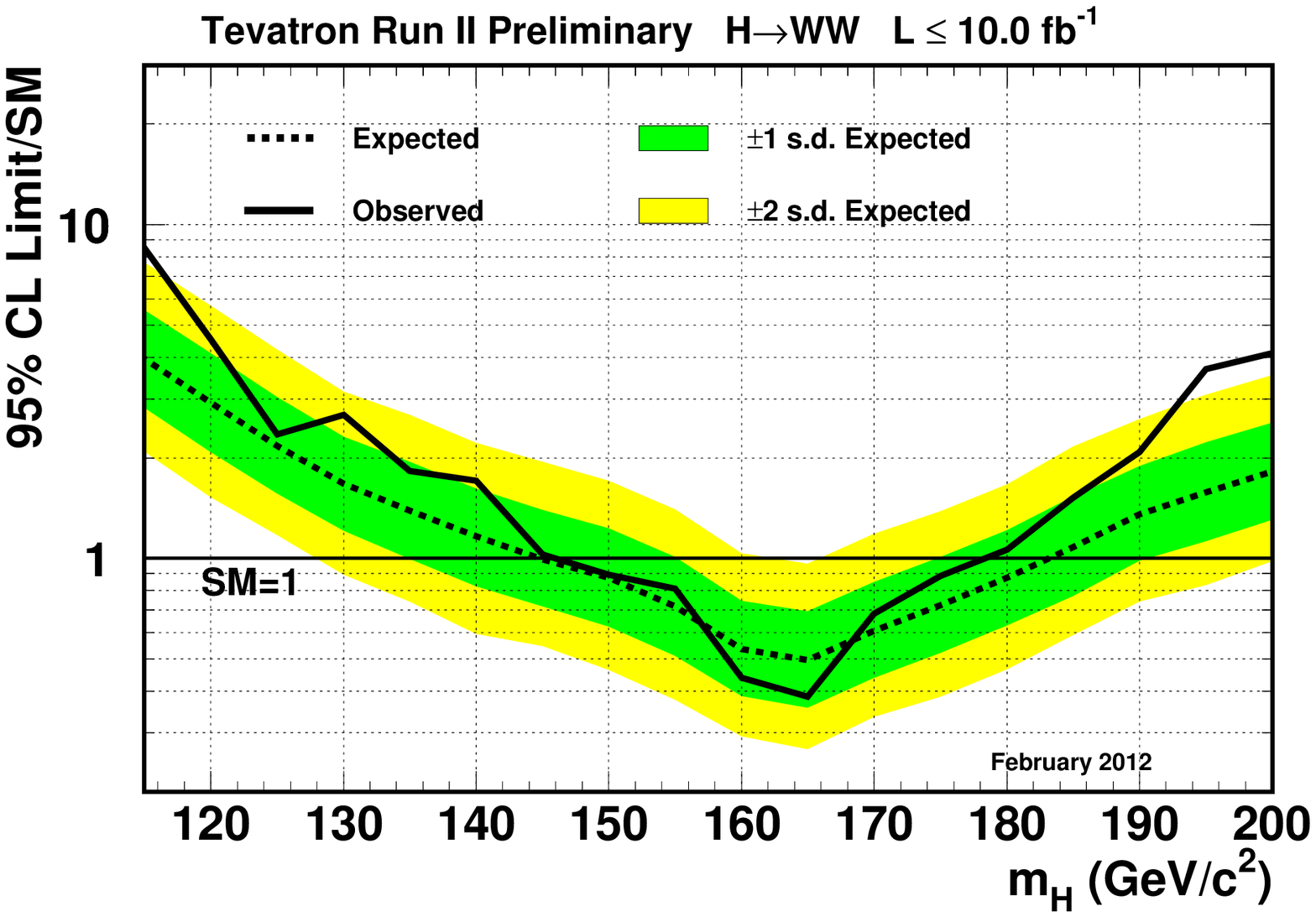,width=6.7cm}}
\vspace*{8pt}
\caption{The Tevatron combined Higgs limit as a function of 
tested Higgs mass in the $H\rightarrow b\bar b$ (left) and 
$H\rightarrow W^+W^-$ (right) decays separately.\label{fig:bbwwlimit}}
\end{figure}

All of the searches for the SM Higgs boson at the Tevatron are combined together
for the best sensitivity~\cite{tevcomb}. We are able to exclude the Higgs mass 
between $100<m_H<106$ and $147<m_H<179$ GeV/c$^2$ with comparable expected 
exclusion $100<m_H<120$ and $141<m_H<184$ GeV/c$^2$, 
as shown in Fig~\ref{fig:tevlimit}. There are some excess of 
events observed in the mass range between 115 and 135 GeV/c$^2$ with a maximum
local p-value of 2.7 standard deviations (sigma) at $m_H$ = 120 GeV/c$^2$, where
the expected local p-value for a SM Higgs signal is 2.0 sigma. When
corrected for the look-elsewhere effect (LEE), which accounts for the 
possibility of selecting the strongest of several random excesses in the 
range $115<m_H<200$ GeV/c$^2$, the global significance of the excess is 2.2
sigma. We also combined results for different decay modes to see where the 
excess comes from. Fig.~\ref{fig:bbwwlimit} shows the combined limit for 
$H\rightarrow b\bar b$ and $H\rightarrow W^+W^-$ separately. 
The observed limit in the $H\rightarrow b\bar b$ channel is more than 2 sigma 
higher than expected in the mass range between 115-135 GeV/c$^2$, 
which counts the majority of the excess in the same mass region. 

\begin{figure}[htpb]
\centerline{\psfig{file=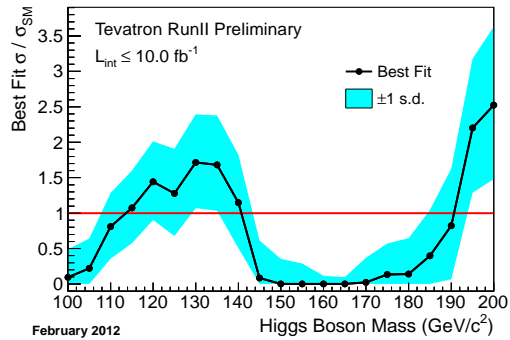,width=6.7cm}}
\vspace*{8pt}
\caption{The fitted signal strength as a function of the Higgs mass, which the
data seem consistent with the SM Higgs signal in the mass range between 
115 and 135 GeV/c$^2$.\label{fig:signalstrength}}
\end{figure}

Given the excess, we fitted the signal production cross section times branching
ratio, normalized to the SM expectation, as a function of Higgs mass, 
as shown in Fig.~\ref{fig:signalstrength}. The obtained signal strength seems 
consistent with the SM Higgs signal in the mass range 
between 115 and 135 GeV/c$^2$. 

\begin{figure}[htpb]
\centerline{\psfig{file=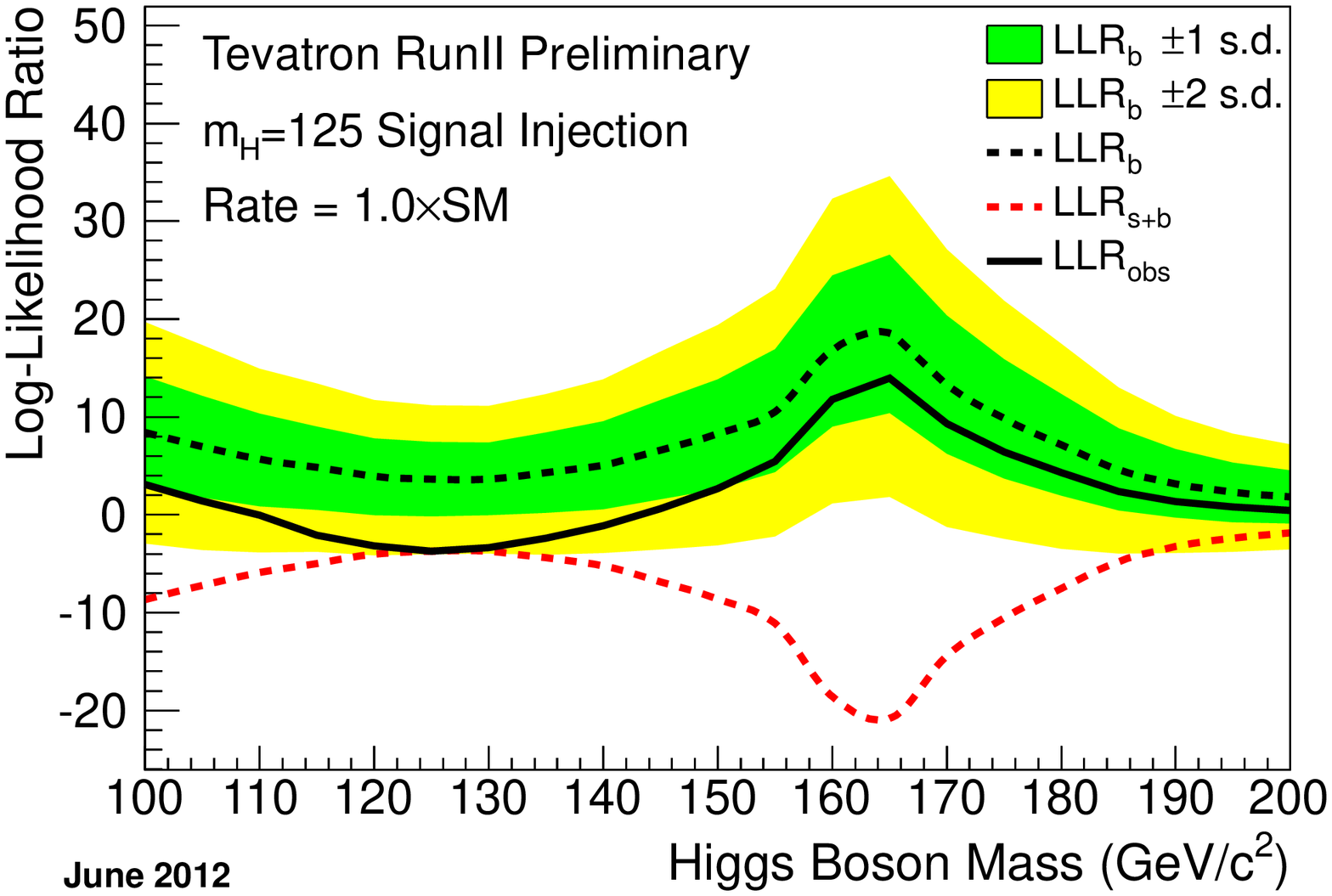,width=6.7cm}}
\vspace*{8pt}
\caption{The expected LLR after injecting an SM Higgs signal at $m_H=125$ 
GeV/c$^2$ into the background-only-pseudo experiment.\label{fig:injecting}}
\end{figure}

To further check the compatibility with SM Higgs signal at $m_H=125$ GeV/c$^2$, 
we compared the LLR by injecting the Higgs signal of 
$m_H=125$ GeV/c$^2$ into the background-only pesudo-experiment (shown in 
Fig.~\ref{fig:injecting}) which seems consistent with the LLR observed in 
the data as shown in Fig.~\ref{fig:LLR}. 
The distribution is quite broad due to the fact that the final
discriminant is not optimized for mass, but for the signal and background  
separation. 
 
Finally, it's worth noting that the Tevatron has made significant progresses to 
achieve the SM Higgs sensitivity. The gain obtained over time seems go 
proportional to the inverse of the extra luminosity used, 
instead of the square root of the luminosity.

\section{Conclusion} 
In clusion, 
with a full dataset and many years of hard work, the Tevatron has 
finally achieved the SM 
Higgs sensitivity over most of the mass range up to 185 GeV/c$^2$.
We observe an excess of events in the data compared with the 
background predictions, which is most significant in the mass range 
between 115 and 135 GeV/c$^2$, consistent with 
the Higgs-like particle recently observed by ATLAS and CMS. The largest local 
significance is 2.7 standard deviations, corresponding to a global significance
of 2.2 standard deviations. We also combine separate searches for 
$H\rightarrow b\bar b$ and $H\rightarrow W^+W^-$, and find that the excess
is concentrated in the $H\rightarrow b\bar b$ channel, although the results 
in the $H\rightarrow W^+W^-$ channel are still consistent with the possible 
presence of a low-mass Higgs boson. 

\section*{Acknowledgment}
We would like to thank the organizers of the phenomenology 2012 symposium for
a wonderful conference with excellent presentations and the CDF and D0 
collaborations for the results presented. A special thanks to Mr. Chee-Hok Lim
for his invitation of this write up. 


\end{document}